\documentclass{article}

\usepackage{amssymb}
\usepackage{color}

\usepackage{algorithm}
\usepackage{ntabbing}
\usepackage{enumerate}
\usepackage{float}

\newfloat{algorithm}{t}{lop}

\def\PF{\noindent{\em Proof:} }

\def\QED{\mbox{}\hspace*{\fill}{$\Box$}\medskip}

\def\bs{\setminus}

\newtheorem{thm}{Theorem}
\newtheorem{lem}[thm]{Lemma}
\newtheorem{propo}[thm]{Proposition}

\def\CC{\mathcal{C}}
\def\Px{$\ell$-L $k$-CP}

\def\Pz{P3C}
\def\LR{$\ell$-L $L$-CP}
\def\IS{$t$-IS}
\def\ISS{$(t - 1)$-IS}
\newcommand{\Oh}{{\mathcal{O}}}
\def\poly{\mbox{poly}}
\def\emptyfunc{f^\emptyset}

\def\ListRecolor{{\sc ListRecolor}}
\def\Recurse{{\sc Recurse}}

\def\restr{|}

\begin{document}

\title{The Complexity of Bounded Length Graph Recoloring}
\author{Paul Bonsma\footnote{Faculty of EEMCS, University of Twente, Enschede, the Netherlands.}
\and Amer E.\ Mouawad\footnote{David R.\ Cheriton School of Computer Science, University of Waterloo, Waterloo, Ontario, Canada.}}

\maketitle

\begin{abstract}
We study the following question: Given are two $k$-colorings $\alpha$ and $\beta$ of a
graph $G$ on $n$ vertices, and integer $\ell$. 
The question is whether $\alpha$ can be modified into $\beta$, by recoloring
vertices one at a time, while maintaining a $k$-coloring throughout, and using at
most $\ell$ such recoloring steps.
This problem is weakly PSPACE-hard for every constant $k\ge 4$.
We show that it is also strongly NP-hard for every
constant $k\ge 4$.
On the positive side,
we give an $\Oh(f(k,\ell) n^{\Oh(1)})$ algorithm for
the problem, for some computable function $f$.
Hence the problem is fixed-parameter tractable when parameterized by $k+\ell$.
Finally, we show that the problem is W[1]-hard (but in XP)
when parameterized only by $\ell$.
\end{abstract}

\section{Introduction}
Given a graph $G$ and a positive integer $k$, the classical NP-complete
$k$-Coloring problem asks for an assignment of at most $k$ colors to the vertices of $G$
such that no two
adjacent vertices receive the same color. Such a color assignment is
called a {\em $k$-coloring} or simply a {\em coloring} of $G$.
Under the {\em reconfiguration framework}, we are interested in structural and
algorithmic questions related to the solution space of
the $k$-Coloring problem. In particular, for any graph $G$ and integer $k$,
we can define the {\em $k$-Color Graph $\CC_k(G)$} as follows. The vertex set
of $\CC_k(G)$ corresponds to all $k$-colorings of $G$ and two colorings
are adjacent if and only if they differ on exactly one vertex.
The integer $k$ is also called the {\em number of admissible colors}.
Given two $k$-colorings of $G$, $\alpha$
and $\beta$, the {\em $k$-Color Path} problem asks if there exists a path
in $\CC_k(G)$ from $\alpha$ to $\beta$.
This is a well-studied problem, which is known to be
solvable in polynomial time for $k\le 3$~\cite{CHJ11}.
In addition, it is PSPACE-complete for every constant $k\ge 4$, even
when restricted to bipartite graphs, and PSPACE-complete for $k=4$ for planar bipartite graphs~\cite{BC09}.
However, it can be solved in polynomial time for 
any $k$, if $G$ is a $(k-2)$-degenerate 
graph~\cite{CHJ08,DFFV06}; see also~\cite{BC09} (planar bipartite graphs are 3-degenerate).

In addition to results on the $k$-Color Path problem, many other results have 
been obtained in the setting of $k$-coloring reconfiguration. For the 
aforementioned PSPACE-complete cases, examples have been explicitly 
constructed where any path from $\alpha$ to $\beta$ has exponential 
length~\cite{BC09}. On the other hand, for the polynomial case $k\le 3$, the 
diameter of components of $\CC_k(G)$ is known to be polynomial~\cite{CHJ11}. 
It has been conjectured that this is also the case for $k$ colors and $k-2$ 
degenerate graphs~\cite{BC09}. This conjecture has been answered positively 
for the subclass of graphs of treewidth $(k-2)$~\cite{BB13}.
The connectedness of $\CC_k(G)$ has been studied in~\cite{CHJ08,CHJ09}. 
In particular, deciding whether $\CC_k(G)$ is connected is coNP-complete 
for $k=3$ and $G$ bipartite, but can be answered in polynomial time if $G$ is planar and bipartite~\cite{CHJ09}.
One may ask whether a {\em shortest} path from $\alpha$ to $\beta$ can be 
found in $\CC_k(G)$. In~\cite{CHJ11}, a polynomial time algorithm for $k=3$ 
and certain pairs of colorings has been given. Finally, it is known that 
for any two $k$-colorings $\alpha$ and $\beta$, $2k-1$ colors suffice to 
find a path from $\alpha$ to $\beta$, and that this bound is tight~\cite{Cer07,MH99,Jac97}.

Similar reconfiguration questions can be formulated for almost any 
search problem, after defining a (symmetric) adjacency relation between solutions. 
Such questions have received considerable attention
in recent literature; see e.g. the survey by Van den Heuvel~\cite{JvdH13}. 
In most cases, the complexity behavior of these problems is similar. 
For instance, deciding whether a path exists between two solutions 
is often PSPACE-hard in general, although polynomial time solvable 
restricted cases can be identified. For PSPACE-hard cases, it is not 
surprising that shortest paths between solutions can have exponential length. 
More surprisingly, for most known polynomial time solvable cases, shortest paths 
between solutions have been shown to have polynomial length.
Results of this kind have for instance been obtained e.g. for the 
reconfiguration of independent sets~\cite{BKW14,KMM12}, vertex covers~\cite{MNR14}, 
shortest paths~\cite{Bon13,Bon12FSTTCS,KMM11}, or boolean assignments~\cite{IDHPSUU11}.

These problems are interesting for a variety of reasons. 
From an algorithmic standpoint, reconfiguration
problems model dynamic situations in which we seek to transform a solution into
a more desirable one, maintaining feasibility during the process 
(see~\cite{Jac97} for such an application of $k$-Color Path). Reconfiguration
also models questions of evolution; it can represent the evolution of a genotype
where only individual mutations are allowed and all genotypes must satisfy a
certain fitness threshold, i.e. be feasible.
Moreover, the study of reconfiguration yields insights into the
structure of the solution space of the underlying problem, crucial for the design of efficient
algorithms. In fact, one of the initial motivations behind such questions was to study
the performance of heuristics \cite{GKMP09} and random sampling methods \cite{CHJ08},
where connectivity and other properties of the solution space play a crucial role.

In many applications of reconfiguration problems, the existence of a 
path between two solutions is irrelevant, if every such path has 
exponential length. So the more important question is in fact: does there exist 
a path between two solutions of length at most $\ell$, for some integer $\ell$?
Such {\em length-bounded} reconfiguration questions have been 
considered e.g. in~\cite{Bon13,Bon14,GKMP09,KMM11,MNR14,MNRSS13}.
In some cases where the existence of paths between solutions can 
be decided efficiently, one can in fact find {\em shortest} paths efficiently~\cite{Bon13,Bon14,GKMP09}.
On the other hand, NP-hard cases have also been identified~\cite{KMM11,MNR14}.
If we wish to obtain a more detailed picture of the complexity of 
length bounded reconfiguration, the setting of parameterized 
complexity~\cite{DF97,FG06} is very useful, where we choose $\ell$ as parameter.
A systematic study of the parameterized complexity of 
reconfiguration problems was initiated by Mouawad et al~\cite{MNRSS13}. 
However, in~\cite{MNRSS13}, only negative results were obtained for 
length-bounded reconfiguration: various problems were identified where 
the problem was not only NP-hard, but also W[1]-hard, when 
parameterized by $\ell$ (or even when parameterized by $k+\ell$, where $k$ is another problem parameter).
In this paper, we give an example of a length bounded 
reconfiguration problem that is NP-hard, but admits an FPT algorithm. 
Another example, namely vertex cover reconfiguration in graphs of bounded degree, was very 
recently obtained by Mouawad et al in~\cite{MNR14}.
\medskip

\noindent
{\em $\ell$-Length $k$-Color Path (\Px):}\\
{\sc Instance:} A graph $G$ on $n$ vertices, nonnegative integers $k$ and $\ell$, and two $k$-colorings $\alpha$ and $\beta$ of $G$.\\
{\sc Question:} Does $\CC_k(G)$ contain a path from $\alpha$ to $\beta$ of length at most $\ell$?

\medskip
In this paper we explore fully how the complexity of the above problem 
depends on the problem parameters $k$ and $\ell$. Firstly, \Px\ is easily 
observed to be PSPACE-hard in general, for $k\ge 4$: Since there are at 
most $k^n$ different $k$-colorings of a graph on $n$ vertices, a path 
from $\alpha$ to $\beta$ exists if and only if there exists one of 
length at most $k^n$. So setting $\ell=k^n$ yields a trivial 
reduction from the PSPACE-hard $k$-Color Path problem to \Px.
Nevertheless, this only establishes {\em weak} PSPACE-hardness, since the 
chosen value of $\ell$ is exponential in the instance size.
In other words, if we require that all integers are encoded in unary,
then this is not a polynomial reduction.
And indeed, the complexity status of the problem changes under
that requirement; in that case, \Px\ is easily observed to be in NP.
(If $(G, k, \ell, \alpha, \beta)$ is a yes-instance, then a path of length 
$\ell$ in $\CC_k(G)$ from $\alpha$ to $\beta$ is a polynomial certificate.)
In Section~\ref{sec:nphard}, we show that \Px\ is in fact NP-complete when 
$\ell$ is encoded in unary, or in other words: it is {\em strongly NP-hard}.
On the positive side, in Section~\ref{sec:fpt}, we show that the 
problem can be solved in time
$\Oh(2^{k(\ell+1)}\cdot \ell^\ell\cdot \poly(n))$.
This establishes 
that \Px\ is {\em fixed parameter tractable (FPT)}, when parameterized by $k+\ell$.
The reader is referred to the excellent books of Downey and
Fellows~\cite{DF97}, Flum and Grohe~\cite{FG06}, and Niedermeier~\cite{N06} for an introduction to parameterized
complexity.
One may ask whether the problem is still FPT 
when only parameterized by $\ell$. We show that this is not 
the case (unless W[1]=FPT), by showing that \Px\ is W[1]-hard when 
only parameterized by $\ell$. We observe however that a 
straightforward branching algorithm can solve the 
problem in time $n^{\Oh(\ell)}$, so in polynomial time for 
any constant $\ell$. In other words, \Px\ is in XP, parameterized by $\ell$.

\begin{table}
\caption{Summary of the problem complexity and results}
\label{table:summary}
\footnotesize
\begin{tabular}{|l|c|c|c|}
\hline
			& $\ell$ as binary variable & $\ell$ as unary variable & $\ell$ as parameter \\
\hline		
$k$ as input variable	& PSPACE-complete	&  NP-complete	& W[1]-hard (*) \\
(unary or binary)	&			&		& in XP (T) \\
\hline				
$k$ as parameter & (para-)PSPACE-complete & (para-)NP-complete	& FPT (*) \\
			     & 		&  & (see also~\cite{JKKPP14})\\
\hline		
$k\ge 4$ as constant	& PSPACE-complete&NP-complete (*)& FPT		  \\
			&  (T), using~\cite{BC09} &			& 	 	\\
\hline		
$k\le 3$ 		& polynomial~\cite{JKKPP14}	& polynomial& polynomial \\
\hline
\end{tabular}
\end{table}

Our results are summarized in Table~\ref{table:summary}. Our main results are marked by (*), and
trivial results are marked with (T). Unmarked results follow immediately from results in an adjacent row or column.
To emphasize the strength of both the negative and positive results, we added rows
for both $k$ constant and $k$ as parameter, even though the complexity is always the same.
The FPT result was also obtained very recently and
independently by Johnson et al~\cite{JKKPP14}, although they use a very different algorithm than ours.
In addition, in~\cite{JKKPP14}, the case $k\le 3$ is considered, and is 
shown to be polynomial (as conjectured by Cereceda et al~\cite{CHJ11}).

\section{Preliminaries}
For general graph theoretic definitions, we refer
the reader to the book of Diestel~\cite{D05}.
Unless otherwise stated, we assume that each graph $G$ is a
simple, undirected graph with vertex set $V(G)$ and
edge set $E(G)$, where $|V(G)| = n$ and $|E(G)| = m$.
The {\em open neighborhood} of a vertex $v$ is denoted by $N_G(v) = \{u \mid uv \in E(G)\}$ and the
{\em closed neighborhood} by $N_G[v] = N_G(v) \cup \{v\}$.
For a set of vertices $S \subseteq V(G)$,
we define $N_G(S) = \{v \not\in S \mid uv \in E(G), u \in S\}$ and $N_G[S] = N_G(S) \cup S$.
We drop the subscript $G$ when clear from context.
The subgraph of $G$ induced by $S$ is denoted by $G[S]$, where
$G[S]$ has vertex set $S$ and edge set $\{uv \in E(G) \mid u, v \in S\}$.
The maximum degree of a graph $G$ is denoted by $\Delta(G)$.
Let $u$ and $v$ be vertices in a graph $G$. A {\em pseudowalk} from $u$ to $v$ of
length $\ell$ is a sequence $w_0,\ldots,w_{\ell}$ of vertices in $G$
with $w_0=u$, $w_{\ell}=v$, such that for
every $i\in \{0,\ldots,\ell-1\}$, either $w_i=w_{i+1}$, or $w_iw_{i+1}\in E(G)$.

A {\em $k$-color assignment} for a graph $G$ is a function
$\alpha:V(G)\to \{1,\ldots,k\}$ of {\em colors} to the
vertices of $G$. It is a {\em $k$-coloring} if there are
no edges $uv\in E(G)$ with $\alpha(u)=\alpha(v)$. On the
other hand, if there exists such an edge $uv$, then
this edge is said to give a {\em color conflict}.
A graph that admits a $k$-coloring is called {\em $k$-colorable}.
The minimum $k$ such that $G$ is $k$-colorable is
called the {\em chromatic number} of $G$, denoted by $\chi(G)$.
For a $k$-coloring $\alpha$ of $G$, the set of
colors {\em used by $\alpha$} is 
$\{\alpha(v) \mid v\in V(G)\}$.
Pseudowalks in $\CC_k(G)$ from $\alpha$ to $\beta$ are also called
{\em $k$-recoloring sequences} from $\alpha$ to $\beta$.
If there exists an integer $k$ such that $\alpha_0,\ldots,\alpha_m$ is a
$k$-recoloring sequence, then this is called a
{\em recoloring sequence} from $\alpha_0$ to $\alpha_m$.
Clearly, if there exists a $k$-recoloring sequence
from $\alpha$ to $\beta$ of length $\ell$, then $\CC_k(G)$
contains a path from $\alpha$ to $\beta$ of length at most $\ell$.

A {\em $k$-color list assignment} for a graph $G$ is a mapping $L$ that assigns a {\em color list} $L(v)\subseteq \{1,\ldots,k\}$ to each vertex $v\in V(G)$. A $k$-coloring $\alpha$ of $G$ is an {\em $L$-coloring} if $\alpha(v)\in L(v)$ for all $v$. By $\CC(G,L)$ we denote the subgraph of $\CC_k(G)$ induced by all $L$-colorings of $G$, and pseudowalks in $\CC(G,L)$ are called $L$-recoloring sequences. The $\ell$-Length $L$-Color Path (\LR) problem asks, given $G,L,\alpha,\beta,\ell$, where $\alpha$ and $\beta$ are $L$-colorings of $G$, whether there exists an $L$-recoloring sequence from $\alpha$ to $\beta$ of length at most $\ell$. 

For a positive integer $k \geq 1$, we denote $[k] = \{1, \ldots , k\}$.
For a function $f:D\to I$ and subset $D'\subseteq D$, we
denote by $f\restr_{D'}$ the restriction of $f$ to the domain $D'$.
The (unique) trivial function with empty domain is denoted by $\emptyfunc$. Note that for any function $g$, $g\restr_{\emptyset}=\emptyfunc$.
We use $\poly(x_1,\ldots,x_p)$ to denote a polynomial function on variables $x_1,\ldots,x_p$.

\section{W[1]-hardness}\label{sec:whard}
We give a reduction from the following problem:

\medskip
\noindent
{\em $t$-Independent Set (\IS):}\\
{\sc Instance:} A graph $G$ and a positive integer $t$.\\
{\sc Question:} Does $G$ have an independent set of size at least $t$?
\medskip

The \IS\ problem is known to be W[1]-hard~\cite{DF97,FG06} when parameterized by $t$.
We will also use the following result, which was shown
independently by Cereceda~\cite{Cer07}, Marcotte and Hansen~\cite{MH99}
and Jacob~\cite{Jac97}: For every pair of $k$-colorings $\alpha$ and $\beta$ of a graph $G$, there exists a path from $\alpha$ to $\beta$ in $\CC_{2k-1}(G)$, and there are examples where at least $2k-1$ colors are necessary.
The graphs constructed in~\cite{Cer07,MH99,Jac97} to prove the
latter result are in fact very similar. We will use these graphs
for our reduction, though we will need a slightly
stronger claim, so we need to restate the proof.
The next definition and proof
follow~\cite[Section~6.1]{Cer07} closely.

For any integer $k \ge 1$, let $B_k= \overline{K_k \times K_k}$ (the complement of
the Cartesian product graph of two complete graphs on $k$ vertices).
More precisely, we let $V(B_k)=\{b^i_j \mid i,j\in\{1,\ldots,k\}\}$, and two vertices $b^i_j$
and $b^{i'}_{j'}$ are adjacent if and only if $i\not=i'$ and $j\not=j'$. Define
two $k$-colorings $\alpha^k$ and $\beta^k$ for $B_k$ by setting
$\alpha^k(b^i_j)=i$ and $\beta^k(b^i_j)=j$ for all vertices $b^i_j$.

\begin{thm}
\label{thm:NeedsManyColors}
For every integer $k\ge 1$, let $B_k$, $\alpha^k$ and $\beta^k$ be as defined above. Then
\begin{enumerate}
 \item
 every recoloring sequence
 from $\alpha^k$ to $\beta^k$ contains a coloring
 that uses at least $2k-1$ different colors, and
 \item there exists a $(2k-1)$-recoloring sequence of length at most $2k^2$ from $\alpha^k$ to $\beta^k$.
\end{enumerate}
\end{thm}

\PF
Consider a recoloring sequence $\gamma_0,\ldots,\gamma_m$ from $\alpha^k$ to $\beta^k$.
For $i\in \{1,\ldots,k\}$, the vertex set $R_i=\{b^i_j\mid j\in \{1,\ldots,k\}\}$
is called a {\em row} of $B_k$. For every $i$, the coloring $\alpha^k$ colors the
vertices of row $R_i$ all with the same color, and $\beta^k$ colors them all
differently. So we may choose the lowest index $j$ such that $\gamma_j$
colors all vertices of at least one row $R_i$ differently.
The choice of $j$ guarantees that for $\gamma_j$, for all other
rows, there exists a color that is used for at least two different
vertices in the row, and thus is not used in any other row (this
follows easily from the definition of $B_k$). We conclude that
$\gamma_j$ uses at least $2k-1$ different colors in total,
which proves the first statement.

\medskip
To obtain a $(2k+1)$-recoloring sequence from $\alpha^k$ to $\beta^k$, we
can apply the following general method (which applies in
fact to any two $k$-colorings of any $k$-colorable graph):
Choose an arbitrary $k$-coloring $\gamma$ of $B_k$. For every
vertex $v\in V(B_k)$ with $\gamma(v)\le k-1$, recolor $v$ to
the color $k+\gamma(v)$ (this is a color in $\{k+1,\ldots,2k-1\}$, so
it is not used by $\alpha^k$ or $\beta^k$). Next, recolor all vertices
$v$ to their target color $\beta(v)$, starting with those vertices $v$ with $\gamma(v)=k$.
It can be verified that this way, a $k$-coloring is maintained
throughout, and that every vertex is recolored at most twice.
\QED

We remark that the distance from $\alpha^k$ to $\beta^k$ in $\CC_{2k-1}(B_k)$ is
in fact strictly smaller than $2k^2$, but we do not need to
know or prove the exact distance. We can now state the
reduction from \IS\ to \Px\ that we use to prove W[1]-hardness.

\paragraph{Construction}
For ease of presentation, we give a reduction from the \ISS\ problem, which
remains W[1]-hard. Given an instance $(G, t - 1)$ of \ISS,
where $G = (V, E)$ and $V = \{v_1, \ldots, v_n\}$, we construct a
graph $G'$ in time polynomial in $n + m + t$ as follows.

$G'$ contains a copy of $G$ and a copy of $B_{t}$ with all edges between them.
In addition, $G'$ contains $n + t + 1$ independent sets
$C_1$, $\ldots$, $C_{n + t + 1}$, each of size $2t + 2t^2$. We say that
$C_i$ (for $1 \leq i \leq n + t + 1$) is a {\em color-guard set}, as it will be
used to enforce some coloring constraints: in the colorings we define, and all colorings reachable from them using at most $|C_i|-1$ recolorings, $C_i$ will contain at least one vertex of color $i$.
We let $V_G = \{g_1, \ldots, g_n\}$, $V_B = \{b^i_j \mid i,j \in \{1, \ldots, t\}\}$,
$V_C = C_1 \cup \ldots \cup C_{n + t + 1}$, and hence $V(G') = V_G \cup V_B \cup V_C$.
The total number of vertices in $G'$ is therefore $n + t^2 + (n + t + 1)(2t + 2t^2)$.
For every vertex $g_i \in V_G$, we add all edges
between $g_i$ and the vertices in $V_C \setminus (C_i \cup C_{n + t + 1})$.
Similarly, for every vertex $b \in V_B$, we add all edges between $b$
and the vertices in $C_{n + t + 1}$.
We define $\alpha$ as follows. 
For every vertex $g_i \in V_G$, $1 \leq i \leq n$, we set $\alpha(g_i) = i$.
For every $i\in \{1,\ldots,n + t + 1\}$ and every vertex $c\in C_i$, we set $\alpha(c) = i$.
For every vertex $b^i_j \in V_B$ (with $i,j\in \{1,\ldots,t\}$), we choose $\alpha(b^i_j) = n + i$.

\begin{propo}
\label{prop:colorguard}
Let $G'$ and $\alpha$ be as constructed above, from a graph $G$ on $n$ vertices.
Let $k = n + t + 1$, and
let $\alpha_0, \ldots, \alpha_\ell$ be a $k$-recoloring sequence of length
at most $2t + 2t^2$ with $\alpha_0=\alpha$.
Then for all $g_i \in V_G$ and $0 \leq x \leq \ell$, we have $\alpha_x(g_i) \in \{i, n + t + 1\}$.
Moreover, for all $b \in V_B$ and $0 \leq x \leq \ell$, we have $\alpha_x(b) \neq n + t + 1$.
\end{propo}

\PF
Since every vertex $g_i \in V_G$ is adjacent to all vertices in
$V_C \setminus \{C_i \cup C_{n + t + 1}\}$, and $|C_j| = 2t + 2t^2$ for all $j$,
it follows that if $g_i$ receives a color $j\not\in \{i, n + t + 1\}$, then 
all vertices in $C_j$ must have been recolored earlier, which contradicts the fact that we consider a recoloring sequence of length at most $2t + 2t^2$.
Similarly, every vertex $b \in V_B$ is adjacent to all vertices in $C_{n + t + 1}$.
Therefore, if $b$ receives color $n + t + 1$ then all vertices in $C_{n + t + 1}$ must have been recolored first.
\QED

Finally, we define the target coloring $\beta$: for every vertex $v \in V_G \cup V_C$
we set $\beta(v) = \alpha(v)$. For every
vertex $b^i_j \in V_B$ (with $i,j\in \{1,\ldots,t\}$), we choose $\beta(b^i_j) = n + j$.
In other words, the goal is to change from a `row coloring' to a `column coloring' for $V_B$, while maintaining the same coloring for vertices in $V_G \cup V_C$.
The corresponding \Px\ instance is denoted by 
$(G', k, \ell, \alpha, \beta)$ with $k = n + t + 1$ and $\ell = 2t + 2t^2$.

\begin{lem}
\label{lem:YESstaysYES:W}
Let $G'$, $k$, $\ell$, $\alpha$, and $\beta$ be as
constructed above, from a graph $G$ on $n$ vertices.
If $G$ has an independent set of size at least $t - 1$,
then $\CC_{k}(G')$ contains a path
from $\alpha$ to $\beta$, of length at most $\ell = 2t + 2t^2$.
\end{lem}

\PF
We let $S$ be an independent
set of size $t - 1$ in $G$. The following recoloring sequence is
an $(n + t + 1)$-recoloring sequence from
$\alpha$ to $\beta$ of length at most $2t + 2t^2$:

\begin{itemize}
\item Assign the vertices in $G'$ corresponding to the vertices in $S$ color $n + t + 1$ (one by one).
By our construction of $G'$, none of these steps will introduce any color
conflicts since vertices in $V_G$ are not connected
to vertices in $C_{n + t + 1}$.
\item From Theorem~\ref{thm:NeedsManyColors}, we know that we can
recolor the vertices in $V_B$ using $2t - 1$ colors and at most $2t^2$ recoloring steps.
Since $t - 1$ vertices have been recolored in the previous step, we can now use
the corresponding $t - 1$ colors to apply the $(2t - 1)$-recoloring
sequence to $V_B$.
\item Recolor the $t - 1$ vertices that were initially recolored
by assigning them their original color again (which is also their target color in $\beta$). None of these steps will introduce any color conflicts since at this point
every vertex in $V_B$ is assigned a color greater than $n$ again.
\end{itemize}

Clearly, the described recoloring sequence consists of
at most $2(t - 1) + 2t^2 < 2t + 2t^2$ recoloring steps.
\QED

\begin{lem}
\label{lem:NOstaysNO:W}
Let $G'$, $k$, $\ell$, $\alpha$, and $\beta$ be as
constructed above, from a graph $G$ on $n$ vertices.
If $\CC_{k}(G')$ contains a path from $\alpha$ to $\beta$ of
length at most $\ell = 2t + 2t^2$, then $G$ has an independent set of size at least $t - 1$.
\end{lem}

\PF
Assume that there exists some
$k$-recoloring sequence from $\alpha$ to $\beta$ of length at most $\ell = 2t + 2t^2$.
Theorem~\ref{thm:NeedsManyColors} shows that this sequence contains a coloring $\gamma$ that assigns at least $2t-1$ different colors to $V_B$. 
Let $U=\{\gamma(b) \mid b\in V_B\}$ denote this color set with $|U|\ge 2t-1$. 
Vertices in $V_B$ cannot be assigned color $n+t+1$ in a sequence of this length (Proposition~\ref{prop:colorguard}), so $|U\cap \{1,\ldots,n\}|\ge t-1$.
Since $G'$ contains all edges between $V_B$ and $V_G$, we conclude that all vertices in $S=\{g_i\in V_G \mid i\in U\}$ must have $\gamma(g_i)\not=i$, and thus $\gamma(g_i)=n + t + 1$ (Proposition~\ref{prop:colorguard}). It follows that $S$ is an independent set of $G$ of size at least $t-1$.\QED

Combining Lemmas~\ref{lem:YESstaysYES:W} and~\ref{lem:NOstaysNO:W} with the fact that \ISS\ is W[1]-hard yields:

\begin{thm}
$\ell$-Length $k$-Color Path is W[1]-hard when parameterized by $\ell$.
\end{thm}

\section{NP-hardness}\label{sec:nphard}
In this section we show that the \Px\ problem
is strongly NP-hard for every fixed constant $k \geq 4$.
We give a reduction from the following problem which
was shown to be NP-complete by Garey, Johnson, and Stockmeyer~\cite{GJS76}.

\medskip
\noindent
{\em Planar Graph 3-Colorability (\Pz):}\\
{\sc Instance:} A planar graph $G$.\\
{\sc Question:} Is $G$ 3-colorable?
\medskip

In contrast to the previous section, here we construct an instance of
the \LR\ problem.
An instance $G,L,\alpha,\beta,\ell$ of the \LR\ problem with $L(v)\subseteq [4]$ for all $v$ is easily
transformed to an instance of \Px, by adding one complete
graph on four vertices $x_i$ with $i\in [4]$ and $\alpha(x_i)=\beta(x_i)=i$,
and edges $vx_i$ for every vertex $v\in V(G)$ and $i\not\in L(v)$. This yields:

\begin{lem}[\cite{BC09}]
\label{lem:listversion}
For any $k \geq 4$, in polynomial time an \LR\ instance
($G$, $L$, $\ell$, $\alpha$, $\beta$), with color lists $L(v) \subseteq [4]$,
can be transformed into an \Px\ instance ($G'$, $k$, $\ell$, $\alpha'$, $\beta'$), such that the distance
between $\alpha$ and $\beta$ in $\CC(G, L)$ (possibly infinite) is the
same as the distance between $\alpha'$ and $\beta'$ in $\CC_k(G')$.
\end{lem}

We will also make heavy use of the notion of $(a,b)$-forbidding paths
and their properties, which were introduced in~\cite{BC09}.
Informally, these are paths that can be added between any pair 
of vertices $u$ and $v$ (provided that $L(u),L(v)\not=[4]$), that
function as a special type of edge, which only excludes the 
color combination $(a,b)$ for $u$ and $v$ respectively, but allows (recoloring to) any other color combination.
More precisely, for $a, b \in [4]$, a path $P$ with 
color lists $L(x)\subseteq [4]$ for all $x\in V(P)$ and end vertices $u$ and $v$ is an {\em $(a, b)$-forbidding path} if the following two properties hold:
\begin{itemize}
 \item For any combination of colors $x\in L(u)$ and $y\in L(v)$, there 
 exists an $L$-coloring $\gamma$ for $P$ with $\gamma(u)=x$ and 
 $\gamma(v)=y$ if and only if $x\not=a$ or $y\not=b$. Such a pair $(x,y)$ is called {\em admissible} for $P$.
 \item
 For any $L$-coloring $\gamma$ of $P$ and any admissible pair $(x,y)$ with 
 $x=\gamma(u)$ or $y=\gamma(v)$: there exists an $L$-recoloring sequence 
 from $\gamma$ to an $L$-coloring $\delta$ with $\delta(u)=x$ and 
 $\delta(v)=y$, in which every internal vertex is recolored at 
 most once, and $u$ and $v$ are not recolored until the last step.
\end{itemize}
(Note that an $(a,b)$-forbidding path from $u$ to $v$ is
a $(b,a)$-forbidding path from $v$ to $u$.)
The second property implies that such an $L$-recoloring sequence has length at most $|V(P)|-1$.
For instance, a $(1,4)$-forbidding path from $u$ to $v$ with 
$L(u)=\{1,2,3\}$ and $L(v)=\{2,3,4\}$ can be obtained by assigning 
the following lists along a path of length 6 (starting with $L(u)$ and ending 
with $L(v)$): $\{1,2,3\}$, $\{1,4\}$, $\{2,4\}$, $\{2,3\}$, $\{1,3\}$, $\{1,4\}$, $\{2,3,4\}$. 
It can be verified that the two properties above hold. 
In~\cite{BC09}, it was observed that for any 
combination of $L(u)$, $L(v)$ and $(a,b)$ with 
$L(u),L(v)\not=[4]$, an $(a,b)$-forbidding path can be constructed, of length six. 
We repeat the construction here for completeness.

\begin{lem}[\cite{BC09}]
\label{lem:forbidden}
For any $L_u \subset [4]$, $L_v \subset [4]$, $a \in L_u$
and $b \in L_v$, there exists an $(a, b)$-forbidding $(u, v)$-path $P$ of length six
with $L(u) = L_u$, $L(v) = L_v$ and all other color lists $L(w) \subseteq [4]$.
\end{lem}
\PF
We can choose colors $c\in [4]\bs L_u$ and $d\in  [4]\bs L_v$. 
Next, we can choose colors $e\in [4]\bs \{a,c\}$ 
and $f\in [4]\bs \{b,d\}$ with $e\not=f$.
Assign the following colors lists along the path: 
$L_u$, $\{a,c\}$, $\{c,e\}$, $\{e,f\}$, $\{f,d\}$, $\{d,b\}$, $L_v$.
It can be verified that this path satisfies the desired properties.
\QED

The reader is referred to~\cite{BC09} for more details on the
list recoloring version of the problem and the use of $(a,b)$-forbidding paths.

\medskip

\paragraph{Construction}
Given an instance $G$ of \Pz, we construct an instance
($G'$, $L$, $\ell$, $\alpha$, $\beta$) of \LR\ as follows.
Start with the vertex set $V(G)$. All of these vertices
$u\in V(G)$ receive color $\alpha(u)=1$ and $L(u)=\{1,2,3\}$.
For every edge $uv\in E(G)$, add three vertices $x_{uv}$, $y_{uv}$ and $z_{uv}$, with
$\alpha(x_{uv})=4$,
$\alpha(y_{uv})=4$,
$\alpha(z_{uv})=4$,
$L(x_{uv})=\{1,2,4\}$,
$L(y_{uv})=\{3,4\}$, and
$L(z_{uv})=\{1,2,4\}$.
We add the following edges and paths between these vertices.

\begin{itemize}
\item
Add edges $ux_{uv}$ and $uy_{uv}$.
\item
Add a $(1,2)$-forbidding path from $u$ to $x_{uv}$.
\item
Add a $(3,1)$-forbidding path from $u$ to $x_{uv}$.
\item
Add a $(2,3)$-forbidding path from $u$ to $y_{uv}$.
\item
Add a $(2,1)$-forbidding path from $v$ to $x_{uv}$.
\item
Add a $(3,2)$-forbidding path from $v$ to $x_{uv}$.
\item
Add a $(1,3)$-forbidding path from $v$ to $y_{uv}$.
\item
Add a $(4,1)$-forbidding path from $x_{uv}$ to $z_{uv}$.
\item
Add a $(4,2)$-forbidding path from $y_{uv}$ to $z_{uv}$.
\end{itemize}
In all cases, we choose an $(a,b)$-forbidding path of length six (Lemma~\ref{lem:forbidden}).
We denote $Z=\{z_{uv} \mid uv\in E(G)\}$.
Informally, these edges and forbidding paths ensure the following property
(see Lemma~\ref{lem:x_verts_colors} below for details): whenever $u$ and $v$
have the same color, both $x_{uv}$ and $y_{uv}$ must
have color 4, and therefore $z_{uv}$ must have color 4.
On the other hand, when $u$ and $v$ currently have
different colors, then either $x_{uv}$ or $y_{uv}$ can be
recolored to a different color, and thus $z_{uv}$ can either be recolored to $1$ or $2$.

Next, we add four vertices $a,b,c,d$ to $G'$, with the following
colors and lists: $\alpha(a)=1$, $\alpha(b)=2$, $\alpha(c)=3$,
$\alpha(d)=4$, $L(a)=\{1,2,3\}$, $L(b)=\{1,2\}$, $L(c)=\{3,4\}$,
$L(d)=\{4\}$. Add all edges between these vertices, except for $cd$.
Add edges from all vertices in $Z$ to $c$. This yields the graph $G'$.
Note that for all $(a,b)$-forbidding paths that we added, we chose the 
colors $\alpha$ of the end vertices $p$ and $q$ such that $\alpha(p)\not=a$ 
or $\alpha(q)\not=b$. So we can extend $\alpha$ to an $L$-coloring of the 
entire graph (by the definition of $(a,b)$-forbidding paths); we do this in an arbitrary way.

Finally, we define the target coloring $\beta$: for all vertices
$v\in V(G')\bs \{a,b\}$, set $\beta(v)=\alpha(v)$. We set
$\beta(a)=2$ and $\beta(b)=1$, so the goal is to reverse
the colors of these two vertices, while keeping all other colors the same.

\begin{lem}
\label{lem:x_verts_colors}
Let $\gamma$ be an $L$-coloring of $G'$, and let $uv\in E(G)$. If $\gamma(u)=\gamma(v)$, then $\gamma(z_{uv})=4$.
On the other hand, if $\gamma(u)\not=\gamma(v)$, then there exists an $L$-recoloring sequence 
of length at most $\Oh(1)$ from $\gamma$ to an $L$-coloring $\gamma'$ with $\gamma'(z_{uv})\not=4$ and with
$\gamma(w)=\gamma'(w)$ for all $w\in V(G)\cup Z$.
\end{lem}

\PF
We first show that if $\gamma(z_{uv})\not=4$, then $\gamma(u)\not=\gamma(v)$.
Suppose that $\gamma(z_{uv})=1$. Then the $(4,1)$-forbidding path from $x_{uv}$ to $z_{uv}$ ensures that $\gamma(x_{uv})\in \{1,2\}$.
If $\gamma(x_{uv})=1$, then the edge $ux_{uv}$ and $(3,1)$-forbidding path from $u$ ensure that $\gamma(u)\not\in \{1,3\}$, so $\gamma(u)=2$. In addition, the $(2,1)$-forbidding path from $v$ to $x_{uv}$ forces $\gamma(v)\not=2$, so in this case, $\gamma(u)\not=\gamma(v)$. Similarly, if $\gamma(x_{uv})=2$ then considering the indicent edge $ux_{uv}$ and incident forbidding paths, we conclude that $\gamma(u)=3$ and $\gamma(v)\not=3$, so again $\gamma(u)\not=\gamma(v)$.

Next, suppose that $\gamma(z_{uv})=2$. Then the $(4,2)$-forbidding path from $y_{uv}$ to $z_{uv}$ ensures that $\gamma(y_{uv})=3$. Similar to before (considering the remaining forbidding paths), this implies that $\gamma(u)=1$ and $\gamma(v)\not=1$, so $\gamma(u)\not=\gamma(v)$. We have now proved the statement in all cases. Hence if $\gamma(u)=\gamma(v)$, then $\gamma(z_{uv})=4$.

Now suppose $\gamma(u)\not=\gamma(v)$. Then considering the same color combinations and forbidding paths as above, we conclude that in every case, $x_{uv}$ can be recolored to 1 or 2, or $y_{uv}$ can be recolored to 3, possibly after first recoloring some internal vertices of incident forbidding paths, but without recoloring any other vertices. Next, $z_{uv}$ can be recolored to 1 or 2, again after possibly recoloring some internal vertices of the incident forbidding paths. This shows that $z_{uv}$ can be recolored to a color different from 4 in $\Oh(1)$ recoloring steps, without changing the color of any other vertex in $V(G)\cup Z$.
\QED

\begin{lem}
\label{lem:3col_recol}
If $G$ is 3-colorable, then there exists an $L$-recoloring 
sequence for $G'$ from $\alpha$ to $\beta$, of length $\Oh(m)$, where $m=|E(G)|$.
\end{lem}

\PF
The vertices of $G$ form an independent set in $G'$. Therefore, if $G$ is 3-colorable we
can start by recoloring the corresponding vertices in $G'$ to such a 3-coloring, using the colors $1,2,3$.
Since all of the $(a,b)$-forbidding paths in $G'$ from a vertex $u\in V(G)$ 
to another vertex $v$ satisfy $\alpha(v)\not=b$ and $v\not\in V(G)$, it follows 
from the definition that this can be done while recoloring all vertices from 
$V(G)$ and all internal vertices of $(a,b)$-forbidding paths at most once, and recoloring 
no other vertices of $G'$. So this can be done using $\Oh(m)$ recoloring steps.

Now by Lemma~\ref{lem:x_verts_colors}, we can subsequently recolor
the vertices of these $(a,b)$-forbidding paths and all vertices in
$Z$, such that all vertices in $Z$ receive color 1 or 2. For every
vertex in $Z$, this can be done in $\Oh(1)$ recoloring steps, so
this gives $\Oh(m)$ steps in total.

Next, we can recolor $c$ to color 4.
At this point, $a$ can temporarily be recolored to 3, $b$ to $1$, and next $a$ to $2$.
Then we can reverse all previous vertex recolorings, except for $a$ and $b$, and
end up with the coloring $\beta$. The total length of this $L$-recoloring sequence is in $\Oh(m)$.
\QED

\begin{lem}
\label{lem:recol_3col}
If there exists an $L$-recoloring sequence from $\alpha$ to $\beta$ for $G'$, then $G$ is 3-colorable.
\end{lem}

\PF
By considering the vertices $a,b,c,d$, we see that any such 
$L$-recoloring sequence must contain a coloring $\gamma$ with $\gamma(c)=4$.
This implies that for every $z\in Z$, $\gamma(z)\in \{1,2\}$.
Then Lemma~\ref{lem:x_verts_colors} shows that for every $uv \in E(G)$,
$\gamma(u)\not=\gamma(v)$. Hence $\gamma$ restricted to $V(G)$ is a 3-coloring of $G$. 
\QED

Combining Lemmas~\ref{lem:3col_recol} and~\ref{lem:recol_3col} with the fact that we can easily
transform the \LR\ instance to an \Px\
instance (Lemma~\ref{lem:listversion}), and the NP-hardness of \Pz,
shows that \Px\ is {\em strongly} NP-hard. For this it is essential that 
the parameter $\ell$ in the reduction is polynomial in the size of the \Pz-instance. 
More precisely, we have the following result:

\begin{thm}
For any constant $k\ge 4$, the problem \Px, with $\ell$ encoded in unary, is NP-complete.
\end{thm}

\section{FPT algorithm}\label{sec:fpt}

For every constant $\ell$, the \Px\ problem can be solved in polynomial time, using the following simple branching algorithm. Denote the given instance by $(G, k, \ell, \alpha, \beta)$, with $|V(G)|=n$.  
Start with with the intial $k$-coloring $\alpha$. For every coloring generated by the algorithm, consider all possible $k$-colorings that can be obtained from it using one recoloring step. Recurse on these choices, up to a recursion depth of at most $\ell$. Return yes if and only if in one of the recursion branches, the target coloring $\beta$ is obtained. Clearly, this algorithm yields the correct answer. For one coloring, there are at most $kn$ possible recoloring steps, so branching with depth $\ell$ shows that at most $\Oh((kn)^\ell)$ colorings will be considered. This shows that for parameter $\ell$, the problem is in XP. 

Because \Px\ is NP-hard for every constant $k\ge 4$, a similar result cannot be obtained for the parameter $k$, unless $P=NP$. 
Since the \Px\ problem is W[1]-hard when
parameterized by $\ell$, we also do not expect any algorithms solving the problem in $\Oh(h(\ell)(kn)^{\Oh(1)})$ time (for any computable function $h$). 
However, we shall see in this section
that the problem is in fact fixed-parameter tractable
when parameterized by $k + \ell$: it can be solved in
$\Oh(f(k,\ell) n^{\Oh(1)})$ time, for some computable function $f$.

We first give a high-level description of the algorithm.
Denote $S=\{v\in V(G)\mid \alpha(v)\not=\beta(v)\}$.
Clearly, when $|S| > \ell$ we have a no-instance and when $|S| = 0$ we have a trivial yes-instance. In what follows, we assume $0 < |S| \leq \ell$. 
The main challenge that we need to overcome is that the number of vertices that potentially need to be recolored cannot easily be bounded by a function of $\ell$; in particular, there may be too many vertices at distance at most $\ell$ from $S$. However, once we know which vertices will be recolored, the problem can be solved using a branching algorithm similar to the one above.

Let $R=\alpha_0,\ldots,\alpha_{\ell}$ be a recoloring sequence for a graph $G$.
For every vertex $v\in V(G)$, the {\em used-color lists} for $R$ are
defined as $U(v)=\{\alpha_i(v) \mid i\in\{0,\ldots,\ell\}\}$. Note that
a vertex $v$ is recolored at least once in $R$ if and only if $|U(v)|\ge 2$.
In addition we have the following simple but useful observation.

\begin{propo}
\label{propo:limited_options}
Let $R$ be a recoloring sequence for $G$ of length $\ell$, and let $U$ denote
the used-color lists for $R$. Then $\sum_{v\in V(G)} (|U(v)|-1)\le \ell$.
\end{propo}

Our main algorithm to solve the \Px\ problem is Algorithm~\ref{alg:main}, which
uses the subroutine given in Algorithm~\ref{alg:listrecolor}.
The \textsc{Recolor} algorithm (Algorithm~\ref{alg:main}) consists of a two-stage branching algorithm.
The first stage of the algorithm ignores the ordering of recoloring steps and simply tries to ``guess'' the used-color lists for each vertex, assuming that a recoloring sequence $R$ from $\alpha$ to $\beta$ of length at most $\ell$ exists.
These guesses are stored in the lists $L(v)$. 
Clearly, $\{\alpha(v),\beta(v)\}\subseteq L(v)$ should hold.
To construct these lists $L$, the algorithm maintains two disjoint sets of vertices $A$ and $B$ as follows. 
All vertices in $A\cup B$ will be recolored at least once, according to our current guess. Vertices are in $B$ if we have already guessed a used-color list for them.
Initially, we have $A = S$ and $B = \emptyset$.
While $A$ is not empty, we pick a vertex $v \in A$ and branch on
all possible lists $L(v)$.
We then delete $v$ from $A$ and add it to $B$.
Before continuing with the next vertex from $A$, 
we inspect the neighbors of $v$. If there exists $u \in N_G(v)\bs (A\cup B)$ such
that $\alpha(u) \in L(v)$, we add $u$ to $A$ since $u$ must also be
recolored at least once, assuming that $v$ will indeed receive all colors in $L(v)$.

If we reach a state where $\sum_{v \in B} (|L(v)| - 1) > \ell$, then
the current branch is ignored since these lists cannot correspond to used-color lists of a recoloring sequence of length at most $\ell$.
On the other hand, when $A = \emptyset$ and $\sum_{v \in B} (|L(v)| - 1) \leq \ell$,
we have a ``possible solution''. 
That is, we still need to make sure that there
exists a feasible ordering of the recoloring steps that transforms $\alpha$ to $\beta$.
This is handled by the \textsc{ListRecolor} subroutine (Algorithm~\ref{alg:listrecolor}), which is a branching algorithm similar to one sketched in the beginning of this section, although we only assign colors from the lists $L(v)$. Since $\sum_{v \in B} (|L(v)| - 1) \leq \ell$, at any time there are at most $\ell$ different ways to recolor a vertex, according to the lists. So, branching up to a depth of $\ell$, this yields an FPT algorithm for (such instances of) the \LR\ problem, parameterized by $\ell$. Combined with Algorithm~\ref{alg:main}, which generates all relevant guesses for the used-color lists, this yields an FPT algorithm for the \Px\ problem, parameterized by $k+\ell$. We now present the details of these algorithms, starting with the \textsc{ListRecolor} subroutine (Algorithm~\ref{alg:listrecolor}).

\begin{algorithm}
\caption{{\ListRecolor$(G,L,\alpha,\beta,\ell)$}}
\label{alg:listrecolor}
\begin{ntabbing}
\qquad\=\qquad\=\qquad\=\qquad\=\qquad\=\qquad\=\kill

{\bf Input:} A graph $G$, nonnegative integer $\ell$, color lists $L(v)$
for all $v\in V(G)$,\\ and two $L$-colorings $\alpha$ and $\beta$ of $G$.\\
{\bf Output:} ``YES'' if and only if there exists an $L$-recoloring sequence \\
from $\alpha$ to $\beta$ of length at most $\ell$.\\
\\
Return \Recurse$(\alpha,\ell)$\label{la:list1}\\
\\
Subroutine \Recurse$(\gamma,\ell')$:\\
\\
if $\gamma=\beta$ and $\ell'\ge 0$ then return YES\label{la:list2}\\
if $\ell'\le 0$ then return NO\label{la:list3}\\
for all 
$L$-colorings $\delta$ that can be obtained from $\gamma$ by changing the\label{la:list4}\\
\> \> color of a single vertex $x$ to a different color in $L(x)$:\\
\> if \Recurse$(\delta,\ell'-1)$=YES then return YES\label{la:listcsp5}\\
return NO\label{la:listcsp6}
\end{ntabbing}
\end{algorithm}

\begin{lem}
\label{lem:listRecol} 
Let $(G,L,\alpha,\beta,\ell)$ be an input for Algorithm~\ref{alg:listrecolor}, with $n=|V(G)|$, and let $p=\sum_{x\in V(G)} (|L(x)|-1)$.
In time $\Oh(p^\ell \cdot \poly(n))$, Algorithm~\ref{alg:listrecolor} decides whether there exists an $L$-recoloring sequence for $G$ from $\alpha$ to $\beta$, of length at most $\ell$.
\end{lem}
\PF
An easy induction proof shows that a recursive call \Recurse$(\gamma,\ell')$ in Algorithm~\ref{alg:listrecolor} returns YES if and only if there exists an $L$-recoloring sequence from $\gamma$ to $\beta$ of length at most $\ell'$: Line~\ref{la:list4} guarantees that every new $\delta$ that is generated is again an $L$-coloring, which is adjacent to $\gamma$ in $\CC(G,L)$. This shows that the algorithm is correct.

For every $L$-coloring $\gamma$, there are $p$ ways to change the color of some vertex $x\in V(G)$ to a different color in $L(x)$. So at most $p$ new $L$-colorings are generated in one recursive call. Obviously, the recursion depth is at most $\ell$, so this shows that at most $\Oh(p^\ell)$ recursive calls are made in total. One recursive call takes time $\poly(n)$, so this yields the stated complexity bound. 
\QED

\begin{algorithm}
\caption{{\sc Recolor$(G,k,\ell,\alpha,\beta)$}}
\label{alg:main}

\begin{ntabbing}
\qquad\=\qquad\=\qquad\=\qquad\=\qquad\=\qquad\=\kill

{\bf Input:}
A graph $G$, nonnegative integers $k$ and $\ell$, and \\
two $k$-colorings $\alpha$ and $\beta$.\\

{\bf Output:} ``YES'' if and only if there exists a $k$-recoloring sequence \\
from $\alpha$ to $\beta$ of length at most $\ell$.\\
\\

$S:=\{v\in V(G) \mid \alpha(v)\not=\beta(v)\}$\label{la:1}\\
if $|S|>\ell$ then return ``NO''\label{la:2}\\
if $|S|=0$ then return ``YES''\label{la:3}\\
Return \Recurse$(S,\emptyset,\{\emptyfunc\})$\label{la:5}\\
\\
Subroutine \Recurse$(A,B,L)$:\\
\\
if $\sum_{v\in B} (|L(v)|-1)>\ell$ then return ``NO''\label{la:6}\\
if $A=\emptyset$ then return \ListRecolor$(G[B],L,\alpha\restr_B,\beta\restr_B)$\label{la:7}\\
Choose $v\in A$.\label{la:8}\\
$B':=B\cup \{v\}$\label{la:9}\\
for each $U\subseteq [k]$ with $2\le |U|\le \ell$ and $\{\alpha(v),\beta(v)\}\subseteq U$:\label{la:11}\\
\> $L'(v):=U$\label{la:12}\\
\> for each $u\in B$:\label{la:13}\\
\> \> $L'(u):=L(u)$\label{la:14}\\
\> $A':=A\bs \{v\}$\label{la:15}\\
\> for each $u\in N(v)\bs (A\cup B)$ with $\alpha(u)\in U$:\label{la:16}\\
\> \> $A':=A'\cup \{u\}$\label{la:17}\\
\> if \Recurse$(A',B',L')$=``YES'' then return ``YES''\label{la:18}\\
return ``NO''\label{la:19}
\end{ntabbing}
\end{algorithm}

\begin{lem}
\label{lem:invariant}
Let $\alpha$ and $\beta$ be two $k$-colorings for a graph $G$, and $\ell\in \mathbb{N}$.
For every recursive call \Recurse$(A,B,L)$ made by Algorithm~\ref{alg:main} on this input, the following conditions hold:
\begin{enumerate}[(1)]
 \item\label{it:one} $A$ and $B$ are disjoint subsets of $V(G)$.
 \item\label{it:two} For every $u\in V(G)\bs B$: $u\in A$ if and only if
   \begin{enumerate}[(a)]
   \item\label{it:a} $\alpha(u)\not=\beta(u)$, or
   \item\label{it:b} there exists an edge $uv\in E(G)$ with $v\in B$ and $\alpha(u)\in L(v)$.
   \end{enumerate}
\end{enumerate}
\end{lem}

\PF
The first time the subroutine \Recurse\ is called (in Line~\ref{la:5}),
$B=\emptyset$, and $A$ contains exactly those vertices
that have different colors in $\alpha$ and $\beta$, so clearly the above conditions are satisfied.

Now consider a call \Recurse$(A,B,L)$ where the arguments satisfy the given
conditions, and an iteration of the for-loop in Line~\ref{la:11} where a
subsequent call \Recurse$(A',B',L')$ is made. Lines~\ref{la:9},~\ref{la:15} and~\ref{la:17}
show that $A'$ and $B'$ are again disjoint (in Line~\ref{la:17}, only vertices
outside of $B'$ are added), so Condition~(\ref{it:one}) is again satisfied.
For vertices in $A\bs \{v\}$, it still holds that at least one of the
Conditions~(\ref{it:a}) and~(\ref{it:b}) is satisfied (also with respect
to the new lists $L'$). Lines~\ref{la:16} and~\ref{la:17} show that
the newly added vertices in $A'$  are exactly those that now satisfy
Condition~(\ref{it:b}), since $L'(v)=U$ and $v$ is the only new vertex in $B'$.
We conclude that both Conditions~(\ref{it:one}) and~(\ref{it:two}) are maintained for $A',B',L'$.
By induction, it follows that for every recursive call
\Recurse$(A,B,L)$, the above conditions hold for $A$, $B$ and $L$.
\QED

\begin{lem}
\label{lem:main_alg_correct_easy}
Let $\alpha$ and $\beta$ be two $k$-colorings for a graph $G$, and $\ell\in \mathbb{N}$. 
If Algorithm~\ref{alg:main} returns ``YES'' then there exists a 
$k$-recoloring sequence for $G$ from $\alpha$ to $\beta$ of length at most $\ell$.
\end{lem}

\PF
Consider a recursive call \Recurse$(A,B,L)$.
If ``YES'' is returned in Line~\ref{la:7}, then $A=\emptyset$, and there exists an $L$-recoloring sequence for $G[B]$ from $\alpha\restr_B$ to $\beta\restr_B$ of 
length at most $\ell$. 
Since $A=\emptyset$, this also yields a valid recoloring sequence for the entire graph $G$, 
starting from $\alpha$, of the same length. This is because any color 
that is assigned to a vertex $v\in B$ is chosen from $L(v)\subseteq [k]$, and therefore does not conflict with the color $\alpha(w)$ of any vertex 
$w\in V(G)\bs B$ (by Condition~(\ref{it:b}) from Lemma~\ref{lem:invariant}).
In addition, Condition~(\ref{it:a}) shows that the recoloring 
sequence that we obtain for $G$ this way ends with $\beta$. 
We conclude that there exists a $k$-recoloring sequence from $\alpha$ to $\beta$ for $G$, of length at most $\ell$.
If YES is returned by a subsequent recursive call in Line~\ref{la:18}, then the claim follows by induction.
\QED

\begin{lem}
\label{lem:main_alg_correct_hard}
Let $\alpha$ and $\beta$ be two $k$-colorings for a graph $G$, and $\ell\in \mathbb{N}$. 
If there exists a 
$k$-recoloring sequence for $G$ from $\alpha$ to $\beta$ of length at most $\ell$
then Algorithm~\ref{alg:main} returns ``YES''.
\end{lem}

\PF
We prove by induction that for every call \Recurse$(A,B,L)$, ``YES'' is returned if
\begin{enumerate}[(1)]
 \setcounter{enumi}{2}
 \item\label{it:three} there exists a $k$-recoloring sequence $R$ from $\alpha$ 
 to $\beta$ for $G$, of length at most $\ell$, such that for 
 every vertex $v\in B$, $L(v)$ is (exactly) the set of colors used by $R$ for $v$.
\end{enumerate}
Applying this statement to the initial call \Recurse$(S,\emptyset,\{\emptyfunc\})$ in Line~\ref{la:5} proves the lemma.

Suppose Condition~(\ref{it:three})
is satisfied for $A,B,L$. We show that ``YES'' is returned. 
Let $R$ be a corresponding recoloring sequence from 
$\alpha$ to $\beta$, of length at most $\ell$, which uses 
exactly the colors $L(v)$ for each $v\in B$.
First, Proposition~\ref{propo:limited_options} shows 
that ``NO'' is not returned in Line~\ref{la:6}, and the algorithm continues.
If there are no vertices $u\in V(G)\bs B$ that satisfy 
Condition~(\ref{it:a}) or~(\ref{it:b}), then $A=\emptyset$ 
(Lemma~\ref{lem:invariant}), so a call to \ListRecolor\ is made in Line~\ref{la:7}.
Restricting all colorings in $R$ to $B$ yields a $k$-recoloring 
sequence for $G[B]$ from $\alpha\restr_B$ to $\beta\restr_B$ that 
uses exactly the colors in $L$ for each vertex, of length at 
most $\ell$, so in this case, ``YES'' is returned.

In the remaining case, we may assume that $A$ is nonempty, and 
a vertex $v\in A$ is chosen in Line~\ref{la:8}. Let $U$ be 
the set of colors used for $v$ by the sequence $R$.
So $\{\alpha(v),\beta(v)\}\subseteq U$. In addition, we argue 
that $|U(v)|\ge 2$. If $\alpha(v)\not=\beta(v)$, this is obvious. 
Otherwise, by Condition~(\ref{it:two}) from Lemma~\ref{lem:invariant}, $v$ has 
a neighbor $u\in B$ with $\alpha(v)\in L(u)$. Since $R$ uses 
exactly the colors $L(u)$ for $u$, $v$ must be recolored at least once, and thus $|U|\ge 2$.
We conclude that $U$ will be considered in an iteration of the for-loop in Line~\ref{la:11}.
Let $L', A', B'$ be the lists and sets as constructed in this iteration.
We observe that $R$ again satisfies the properties from 
Condition~(\ref{it:three}), also with respect to this $L'$ and $B'$. 
Indeed, $B'=B\cup \{v\}$, and $L'(v)=U$, which is exactly the set of 
colors used by $R$ for $v$. For all other vertices $w\in B'$, $L(w)=L'(w)$. 
So we may use induction to conclude that ``YES'' is returned by the 
call \Recurse$(A',B',L')$, and thus in Line~\ref{la:18}, ``YES'' is returned.
\QED

\begin{lem}
\label{lem:main_alg_runtime}
The \textsc{Recolor} algorithm (Algorithm~\ref{alg:main}) runs in $\Oh(2^{k(\ell+1)}\cdot \ell^\ell\cdot \poly(n))$ time, where $n=|V(G)|$.
\end{lem}

\PF
The search-tree produced by the \textsc{Recolor} algorithm
has depth at most $\ell+1$, since every branch increases the quantity
$\sum_{v\in B} (|L(v)|-1)$ by at least $1$ (Line~\ref{la:11}) and the base case is reached whenever $\sum_{v\in B} (|L(v)|-1) > \ell$ (Line~\ref{la:6}).
The number of sets considered in the for-loop in Line~\ref{la:11} is at most $2^k$, so every node in the search-tree has at most $2^k$ children. We conclude that at most $\Oh(2^{k(\ell+1)})$ recursive calls are made in total. 

We now argue that for every recursive call, we spend at most $\Oh(\ell^\ell \cdot \poly(n))$ time in total. For Line~\ref{la:7}, this follows from Lemma~\ref{lem:listRecol}, noting that whenever this line is reached, $p=\sum_{v\in B} (|L(v)|-1)\le \ell$ holds (Line~\ref{la:6}). All other lines can easily be implemented to run in time $\poly(n)$. (Note that we may assume w.l.o.g.\ that $k\le n+1$.) We attribute the time spent in Lines~\ref{la:12}--\ref{la:18} to the resulting recursive call in Line~\ref{la:18}. 
This shows that the entire complexity can be bounded by $\Oh(2^{k(\ell+1)}\cdot \ell^\ell\cdot \poly(n))$.\QED

Combining Lemmas~\ref{lem:main_alg_correct_easy},~\ref{lem:main_alg_correct_hard} and~\ref{lem:main_alg_runtime}
yields the main result of this section:

\begin{thm}
$\ell$-Length $k$-Color Path is fixed-parameter tractable when parameterized by $k + \ell$.
\end{thm}

\bibliographystyle{plain}
\bibliography{FPTrecol}

\end{document}